\documentclass[12pt]{article}
\usepackage{graphicx}

\usepackage{amsmath,amssymb}
\usepackage{slashed}
\usepackage{subfigure}

\newcommand\pubnumber{CP3-18-23 \\ LPT-Orsay-18-44}
\newcommand\pubdate{\phantom{\today}}

\def\CP3{$^{a}$Centre for Cosmology, Particle Physics and Phenomenology (CP3)\\
Universit\'e catholique de Louvain, 1348 Louvain-la-Neuve, Belgium}
\def\LPT{$^{b}$Laboratoire de Physique Th\'eorique, CNRS, \\
Univ. Paris-Sud, Universit\'e Paris-Saclay, 91405 Orsay, France}
\def\MPI{$^{c}$Max Planck Institut f\"ur Kernphysik, 69117 Heidelberg, Germany}
\def\DESY{$^{d}$Deutsches Elektronensynchrotron (DESY),  22607 Hamburg, Germany}
\def\support{\footnote{Work supported by the the European Union's Horizon 2020 research and innovation programme under the Marie Sklodowska-Curie grant agreement No 750627 and by the Fonds de la Recherche Scientifique - FNRS under Grant n$^{\circ}$ IISN 4.4512.10.}}

\def\Title#1{\begin{center} {\Large #1 } \end{center}}
\def\Author#1{\begin{center}{ \sc #1} \end{center}}
\def\Address#1{\begin{center}{ \it #1} \end{center}}

\newcommand\pubblock{\rightline{\begin{tabular}{l} \pubnumber\\
         \pubdate  \end{tabular}}}
\newenvironment{Abstract}{\begin{quotation}  }{\end{quotation}}
\newenvironment{Presented}{\begin{quotation} \begin{center} 
             PRESENTED AT\end{center}\bigskip 
      \begin{center}\begin{large}}{\end{large}\end{center} \end{quotation}}

\def\beq{\begin{equation}}
\def\eeq#1{\label{#1}\end{equation}}
\def\eeqn{\end{equation}}

\def\beqa{\begin{eqnarray}}
\def\eeqa#1{\label{#1}\end{eqnarray}}
\def\eeqan{\end{eqnarray}}

\let\bar=\overbar

\def\Dslash{\not{\hbox{\kern-4pt $D$}}}
\def\dslash{\not{\hbox{\kern-2pt $\del$}}}

\def\ee{e^+e^-}

\def\msb{{\bar{\ssstyle M \kern -1pt S}}}

\def\be{\begin{equation}}
\def\ee{\end{equation}}

\begin{document}
\begin{titlepage}
\pubblock

\vfill
\Title{Leptogenesis, dark matter and neutrino masses}
\vfill
\Author{Michele Lucente$^{a,}$\support, Asmaa Abada$^{b}$, Giorgio Arcadi$^{c}$ and Valerie Domcke$^{d}$}
\Address{\CP3\\ \vspace{0.1\baselineskip} \LPT \\ \vspace{0.1\baselineskip} \MPI \\ \vspace{0.1\baselineskip} \DESY}
\vfill
\begin{Abstract}
We review the viability of the sterile neutrino hypothesis in accounting for three observational problems of the Standard Model of particle physics: neutrino masses and lepton mixing, dark matter and the baryon asymmetry of the Universe. We present two alternative scenarios for the implementation of the sterile fermion hypothesis: the $\nu$MSM and the Inverse Seesaw.
\end{Abstract}
\vfill
\begin{Presented}

NuPhys2017, Prospects in Neutrino Physics

Barbican Centre, London, UK,  December 20--22, 2017

\end{Presented}
\vfill
\end{titlepage}
\def\thefootnote{\fnsymbol{footnote}}
\setcounter{footnote}{0}

\section{Introduction}
The Standard Model (SM) of particle physics provides a coherent and successful framework to account for an incredibly wide set of data. However, there are at least three firm observations that cannot be accounted for in the SM, namely: the fact that neutrinos are massive and leptons mix, the dark matter (DM) component of the Universe and the observed baryon asymmetry of the Universe (BAU). It would be remarkable if all the aforementioned observational problems could be accounted for by a simple (and natural) extension of the minimal SM: the introduction of heavy-neutral leptons\footnote{In the following we refer to HNL that mix with the SM active neutrinos as sterile neutrinos.} (HNL). HNL are absent in the minimal SM, but they arise in the form of right-handed neutrinos (RH$\nu$) by requiring each SM field to exist for both chirality states.

\section{The minimal framework: Type-I Seesaw}
By extending the SM field content by a number $n$ of gauge singlet fermions $N_I$, the Lagrangian of the model gets extended by the following renormalizable interactions
\be\label{eq:Seesaw}
\mathcal{L} = \mathcal{L}_{\text{SM}} + i \bar{N_I} \slashed{\partial} N_I - \left( Y_{\alpha I} \bar{\ell_\alpha} \tilde{\phi} N_I + \frac{M_{IJ}}{2} \bar{N^c_I} N_J + h.c. \right),
\ee
where $\ell_\alpha$ are the SM lepton doublets, $\tilde{\phi} = i \sigma_2 \phi^*$ with $\phi$ the Higgs doublet, $Y_{\alpha I}$ are dimensionless Yukawa couplings, $M_{IJ}$ is a symmetric Majorana mass matrix and the indices run on $\alpha = e,\mu,\tau$ and $I = 1,\dots,n$. The origin of the mass matrix $M$ is unknown: following a bottom-up approach its energy scale must be phenomenologically identified. After the electroweak symmetry breaking (EWSB), the new interactions in eq.~(\ref{eq:Seesaw}) generate a non-zero mass matrix for the active neutrinos which, under the assumption $|Y v| \ll |M|$ ($v = 246$ GeV is the value of the Higgs vacuum expectation value) reads, in the flavour basis,
\be\label{eq:nu_Seesaw}
m_\nu \simeq - \frac{v^2}{2} Y^* M^{-1} Y^\dagger \lesssim 1 \text{ eV}.
\ee
This construction results in the well-known Seesaw mechanism for the generation of neutrino masses. It is remarkable that the very same Lagrangian in eq.~(\ref{eq:Seesaw}) provides, without further assumptions, the ingredients for a viable leptogenesis scenario: the complex Yukawa couplings $Y$ provide, in general, the CP-violating phases, while the new fermion singlets $N_I$ deviate from thermal equilibrium at some time during the early Universe expansion. Finally, the SM sphalerons violate the total baryon number $B$ and lepton number $L$ by rapidly erasing any $B+L$ asymmetry (while preserving any existing $B-L$ charge) as long as they are in thermal equilibrium, for temperatures $T$ such that $10^{12} \text{ GeV} \gtrsim  T \gtrsim T_\text{EW}$, where $T_\text{EW} \simeq 140$ GeV is the temperature of the electroweak phase transition.
In addition, HNL are natural DM candidates as well: they are massive, weakly interacting and, depending on their masses and couplings, they can be metastable on cosmological timescales.

\subsection{Leptogenesis realisations}
The third Sakharov condition states a deviation from thermal equilibrium in the early Universe as a necessary condition to generate a baryon asymmetry: depending on the temperature at which sterile neutrinos deviate from thermal equilibrium, it is possible to classify two main frameworks for leptogenesis. In the first one, usually dubbed thermal leptogenesis~\cite{Fukugita:1986hr}, the size of the Yukawa couplings is large enough such that an equilibrium population of sterile neutrinos is generated shortly after reheating. When the Universe cools down to temperatures below the sterile neutrino masses, the equilibrium number densities of the particles become exponentially suppressed, and if sterile neutrino couplings are sufficiently weak, the actual populations are not able to follow the equilibrium abundance; being unstable particles sterile neutrinos eventually decay out of thermal equilibrium. Due to their Majorana character, the out-of-equilibrium decay of the sterile neutrinos can produce a non vanishing lepton asymmetry $L$, which is then converted into a baryon asymmetry by sphaleron processes.
Thermal (high-scale) leptogenesis can simultaneously account for neutrino physics and for the observed BAU, provided a lower bound on the sterile neutrino mass scale is fulfilled, {$M \gtrsim 10^{8}$ GeV}, for the case of a non-degenerate sterile neutrino mass spectrum~\cite{ref:non_res_lepto}. This lower bound can be relaxed to the TeV scale for a degenerate mass spectrum (resonant leptogenesis), resulting in the condition $M \gtrsim 100$ GeV  if motivated flavour patterns are considered as well~\cite{ref:res_lepto}. Testing the latter mass scales is challenging in current experiments.

An alternative leptogenesis realisation at low scale was proposed by Akhmedov, Rubakov and Smirnov (ARS mechanism)~\cite{Akhmedov:1998qx}. In this scenario the sterile neutrinos are assumed to enter thermal equilibrium at much later times, typically close to the electroweak temperature $T_\text{EW}$: the deviation from thermal equilibrium is thus provided during their production, rather than during their decay. This requirement translates into a condition on the Yukawa couplings $|Y| \lesssim 10^{-6}$ and, recalling eq.~(\ref{eq:nu_Seesaw}), to
\be
m_\nu \simeq - \frac{v^2}{2} Y^* M^{-1} Y^\dagger \simeq 0.3 \left(\frac{\text{GeV}}{M}\right)\left(\frac{Y^2}{10^{-14}}\right) \text{ eV}.
\ee
It is evident that, in order to reproduce the observed neutrino masses, the sterile neutrinos are much lighter than in the thermal leptogenesis scenario, lying at the GeV scale: the ARS mechanism has thus the attractive feature of being testable in current and future experimental facilities.
The generation of the baryon asymmetry in the ARS scenario relies on a different mechanism with respect to the thermal leptogenesis case~\cite{ref:nuMSM,ref:ARS}: given that the neutrino mass scale is much smaller than the plasma temperature $T$, $M \sim \text{GeV} \ll T_\text{EW} \lesssim T$, the neutrino Majorana character is suppressed, and the total lepton number (defined including all active and sterile flavours) is approximately conserved in sterile neutrino interactions (although this is not the case for the whole parameter space~\cite{ref:L-violating}). However, due to the CP-violating nature of the Yukawa couplings $Y$, asymmetries in the individual (active and sterile) lepton flavours arise during the sterile neutrino production, while their sum approximately vanishes: since SM sphalerons only couple to the active leptons, they convert the asymmetry in this sector (and only this asymmetry) into a net baryon asymmetry. Moreover, the final asymmetry is boosted if the sterile neutrinos exhibit a degenerate mass spectrum, since this enhances CP-violating oscillations among different flavours: indeed a degenerate mass spectrum is a necessary condition to reproduce the observed BAU in the case where only 2 RH neutrinos are present, while a non-degenerate mass spectrum is a viable scenario if at least 3 RH neutrinos contribute to the generation of the asymmetry~\cite{Drewes:2012ma}.

\subsection{Sterile neutrinos as dark matter}
Sterile neutrinos (and HNL in general) are in principle viable DM candidates: they are produced in the early Universe by the oscillations of the active neutrinos in thermal equilibrium, as long as an active-sterile mixing is present (Dodelson-Widrow mechanism, DW)~\cite{Dodelson:1993je}. There exist of course a number of observational constraints that limit the available parameter space of a sterile neutrino dark matter, including the ones on the abundance, phase-space density, lifetime (from indirect detection) and structure formation~\cite{Abada:2014zra}. The latter one is especially constraining, but it is also the most model-dependent one: sterile neutrinos produced via the DW mechanism can be classified as warm dark matter, and are subject to strong constraints from the Lyman-$\alpha$ forest data. Combined together, the mentioned constraints restrict the DM neutrino mass at the keV scale, and exclude the viability of a sterile neutrino produced via DW as the dominant DM component, limiting its relative abundance to at most $\sim 30\%$ of the total DM abundance~\cite{Abada:2014zra}. It should however be stressed that the large scale structure formation depends on the DM free-streaming length, and thus on its production mechanism: alternatives to the DW mechanism giving rise to a colder DM momentum distribution (in agreement with observation) are presented in the following.

\section{The $\nu$MSM}
The $\nu$ Minimal Standard Model ($\nu$MSM)~\cite{ref:nuMSM} is a realisation of the Type-I Seesaw featuring a phenomenologically motivated mass spectrum: in this model three right-handed neutrinos are added to the SM field content, two of which ($N_{2,3}$) have degenerate masses at the GeV scale and are at the origin of both the baryon asymmetry of the Universe and neutrino masses, while the third state $N_1$ has mass at the keV scale and does not significantly contribute to the generation of neutrino masses.

In the $\nu$MSM the population of sterile neutrinos is vanishing at large temperatures $T \gg T_\text{EW}$, and the $N_{2,3}$ states approach thermal equilibrium as the Universe expands: the ARS mechanism is at play in this phase, producing an asymmetry in the sterile and active lepton flavours, resulting in a net baryon asymmetry after sphaleron effects are taken into account. The $N_{2,3}$ states eventually thermalise during or after the electroweak phase transition, when the SM sphalerons are not effective and the BAU has frozen-out: during this phase the previously existing lepton asymmetry is washed-out, until the moment when $N_{2,3}$ kinematically freeze-out and then decay out of thermal equilibrium. The decay process is conceptually similar to the thermal leptogenesis scenario, but it happens at much later times, $T\sim$ GeV, when the temperature drops below the (heavy) sterile neutrino masses, and thus the resulting lepton asymmetry is not converted into a baryon asymmetry by the ineffective sphaleron transitions. On the other hand this lepton asymmetry plays an important role in the subsequent DM production: in the presence of a lepton-asymmetric background, the effective potential that drives the conversion of active into sterile neutrinos in the DW mechanism gets modified, similarly to what happens to the vacuum neutrino oscillation parameters in matter (MSW effect). This mechanism, known as Shi-Fuller (SF)~\cite{Shi:1998km}, results in a  lepton number-driven resonant conversion of active into sterile neutrinos, which is peaked at lower momenta with respect to a thermal spectrum. On the one hand, this enhances the active-sterile conversion, requiring smaller active-sterile mixings for the production of the observed DM relic density with respect to the DW mechanism, thus complying with bounds from stability and indirect detection; on the other hand, the sterile neutrinos momentum distribution is ``colder'' with respect to the DW mechanism (where they inherit a thermal spectrum from their active siblings) thus relaxing the bounds from structure formation.
All these ingredients make it possible in the $\nu$MSM to provide simultaneous viable solutions for the three aforementioned observational problems of the SM: as long as neutrino masses and BAU are considered, viable solutions require a relative mass degeneracy in the heavy neutrino pair $N_{2,3}$ of the order $\delta M/M \lesssim 10^{-3}$. Beside, and in order to account for viable DM production as well, a large lepton asymmetry, of about 5 orders of magnitude bigger than the observed BAU, is required in the realisation of the SF mechanism: the generation of this lepton asymmetry in the late-time out-of-equilibrium decay of the heavy states $N_{2,3}$ requires a much stronger mass degeneracy, of the order $\delta M/M \lesssim 10^{-14}$.

\section{The Inverse Seesaw}
The Inverse Seesaw (ISS)~\cite{ref:ISS} is a neutrino mass generation mechanism based on symmetry arguments: it consists in enlarging the SM field content by the addition of a number $\# \nu_R$ of right-handed neutrino fields $\nu_R$ and of further $\# s$ fermionic sterile singlets $s$ with the same lepton number, the difference between them being the fact that the $s$ fields do not feature a Yukawa coupling with the left-handed neutrinos $\nu_L$. In the basis $n_L = (\nu_L, \nu_R^c, s)^T$ the neutrino mass terms read, in the ISS,
\be\label{eq:ISS_lagr}
-\mathcal{L}_{m_\nu} = \frac{1}{2} n_L^T C \mathcal{M} n_L + h.c.,\hspace{0.5cm} \text{ with } \hspace{0.5cm} \mathcal{M} = \left(
\begin{array}{ccc}
0 & d & 0 \\
d^T &0 & n\\
0 & n^T & \mu 
\end{array}
\right),
\ee
where the Dirac mass matrix $d$ is generated after the EWSB, $d=v Y^* / \sqrt{2}$, $n$ is a mass matrix coupling the new fields $\nu_R$ and $s$, $\mu$ is a symmetric Majorana mass matrix for the $s$ fields and $C=i\gamma^2\gamma^0$ is the charge conjugation matrix. The parameter $\mu$ in eq.~(\ref{eq:ISS_lagr}) is the only one that violates the total lepton number $L$: following 't Hooft naturalness argument one can assume that $\mu$ is small compared to the other mass parameters, since in the limit $\mu \rightarrow 0$ the Lagrangian increases its symmetries. After diagonalization, the active neutrino mass matrix is given, in the (seesaw) limit $|\mu| \ll |d| \ll |n|$, by
\be\label{eq:mnu_ISS}
m_\nu \simeq d \left(n^{-1}\right)^T \mu \left(n^{-1}\right) d^T.
\ee
In the ISS it is thus possible to link the smallness of neutrino masses with the smallness of the lepton number violating parameter $\mu$, thus allowing for viable phenomenology even with sizeable Yukawa couplings and a relatively low new physics scale.

The ISS mass spectrum depends on the number of new fields that are introduced~\cite{Abada:2014vea}: it features in general $\# \nu_L=3$ light active neutrinos with masses at the $m_\nu$ scale (\ref{eq:mnu_ISS}), and $2 \# \nu_R$ heavy states (that couple to form $\# \nu_R$ pseudo-Dirac pairs) with masses at the scale $n$ and mass splittings of order $\mu$. Finally, only in the scenario where $\# s > \# \nu_R$,  $(\# s - \# \nu_R)$ light sterile states are present at the $\mu$ scale. These light states can provide a solution to the short-baseline (anti-)neutrino oscillation anomalies (if $\mu\sim$ eV ) or can be viable DM candidates (for $\mu\sim$ keV).
It is possible to identify two minimal ISS realisations: the most minimal one, dubbed (2,2) ISS, consists in the addition of 2 right-handed neutrino fields $\nu_R$ and 2 sterile fields $s$, and features 2 heavy pseudo-Dirac pairs of sterile neutrinos. The next-to-minimal model, the (2,3) ISS, contains 3 sterile fields $s$, resulting in addition in the presence of a light massive sterile state at the scale $\mu$.

\begin{figure}[htb]
\centering
\subfigure{%
   \includegraphics[width=0.49\textwidth]{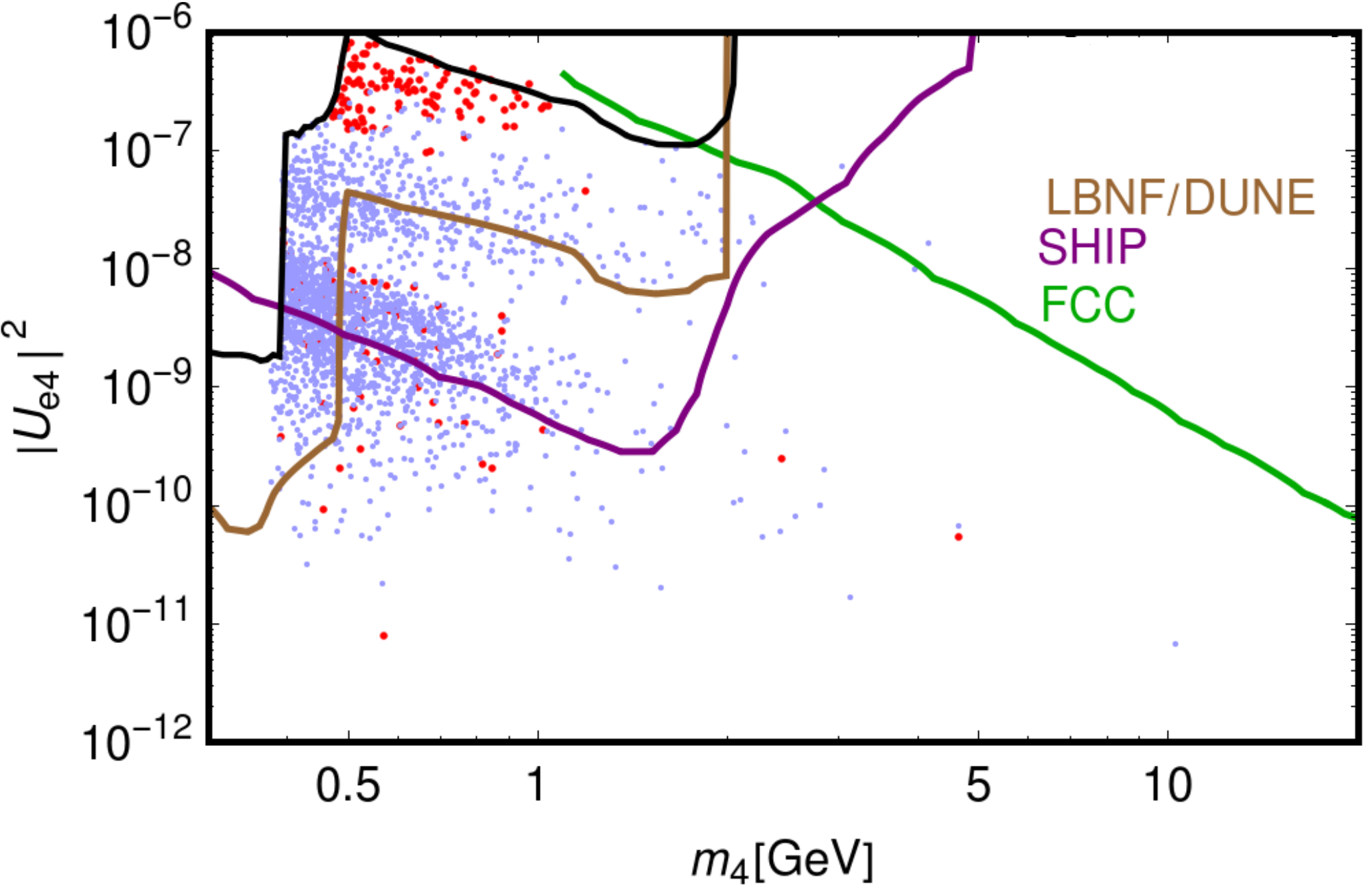}
}\hfill
\subfigure{%
   \includegraphics[width=0.49\textwidth]{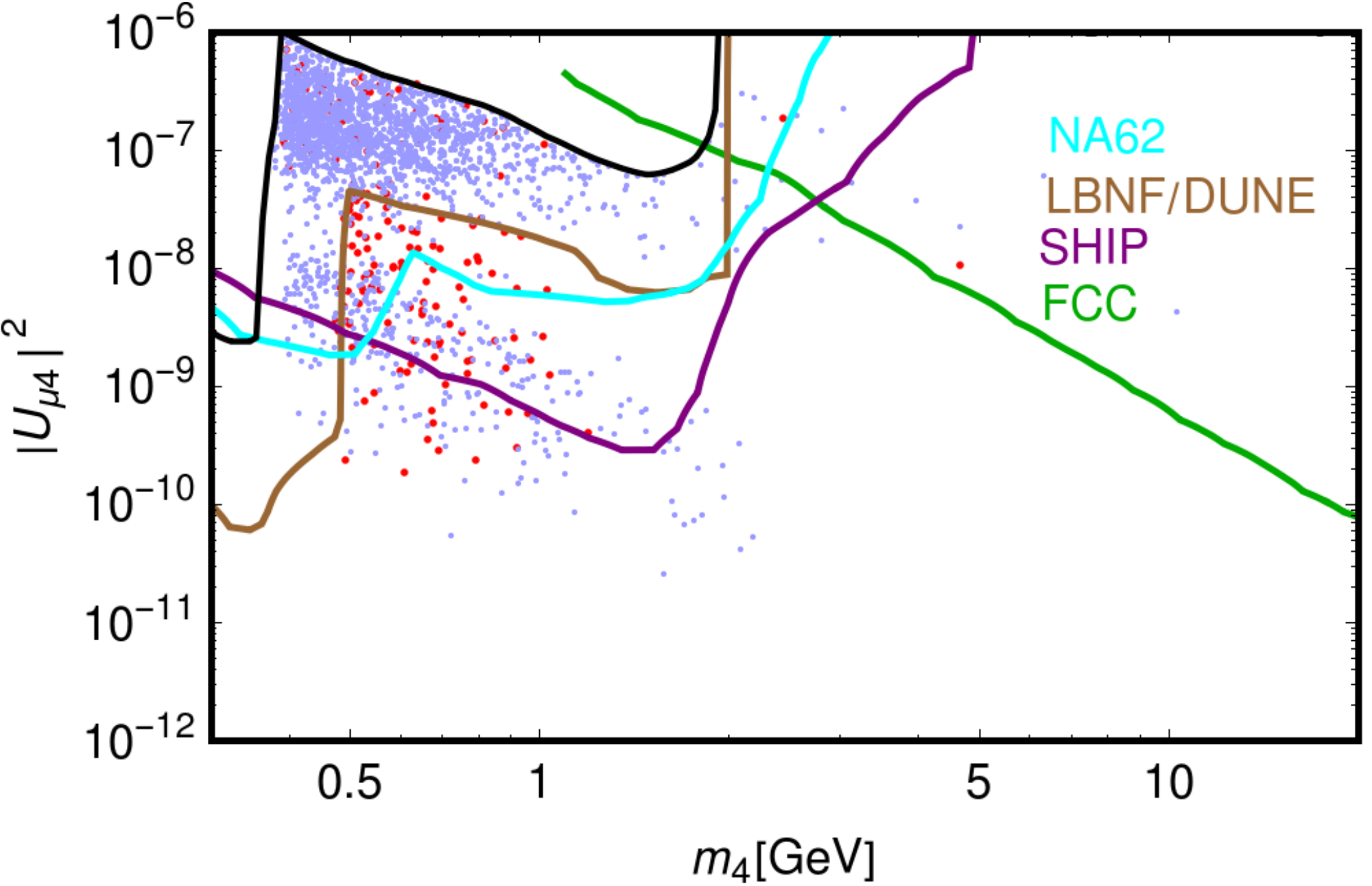} 
}
\caption{Mixing between the active and sterile neutrinos in the electron (left panel) and muon (right panel) flavours, for viable leptogenesis solutions in the (2,2) ISS. The black line denotes existing bounds, while the coloured lines refer to the sensitivity curves of the NA62, LBNF/DUNE, FCC-ee and SHiP experiments. Blue (red) points assume a normal (inverted) hierarchy for the light active neutrinos.}
\label{fig:ISS22_mixing}
\end{figure}

The natural (quasi) degeneracy in the mass spectrum of the ISS allows for an effective leptogenesis at low scales via the ARS mechanism~\cite{Abada:2017ieq}: the mechanism can indeed simultaneously account for viable leptogenesis and neutrino masses with a relative mass degeneracy of order $\delta M/M \lesssim 10^{-2}$ and sterile neutrino masses at the GeV scale. Interestingly, as reported in Figure~\ref{fig:ISS22_mixing}, a large fraction of solutions is testable in current and future experiments such as NA62, LBNF/DUNE and SHiP.
On the other hand, the ISS can simultaneously account for neutrino masses and DM as well~\cite{Abada:2014zra}: this is achieved in the (2,3) ISS realisation, with $\mu$ at the keV scale and $n$ at the TeV scale. In this scenario the Yukawa couplings are large enough such that the heavy pseudo-Dirac neutrinos can thermalise in the early Universe, while the light sterile state does not; this enables the freeze-in production of DM, via the decay of a pseudo-Dirac state into a light sterile state (DM candidate) plus a Higgs boson, the resulting DM abundance being 
\be
\label{eq:heavy_fimp}
\Omega_{\rm DM} h^2 \approx 2 \times 10^{-1} {\left(\frac{\sin\theta}{10^{-6}}\right)}^2 {\left(\frac{m_{\rm s}}{ \text{ keV}}\right)} \sum_I g_I {\left(\frac{Y_{\rm eff,I}}{0.1}\right)}^2 {\left(\frac{\mbox{TeV}}{m_I}\right)} \left(1-\frac{m_h^2}{m_I^2}\right)\varepsilon\left(m_I\right),
\ee
where $\theta$ is the zero temperature mixing between the active neutrinos and the light sterile state, $m_s$ and $m_h$ are the light sterile and Higgs masses, respectively, the index $I$ runs over the pseudo-Dirac states with $g_I$ being their internal degrees of freedom, $Y_{\text{eff,}I}$ their Yukawa couplings in the mass basis, $m_I$ their masses and $\varepsilon(m_I) \in [0,1]$ a function accounting for the temperature dependence of the active-sterile mixing due to the evolution of the Higgs vacuum expectation value. For $m_I\gtrsim 2$ TeV, one has $\varepsilon(m_I)\ll 1$, while for $m_I < m_h$, the decay channel is not kinematically open, resulting in the viable range $n\approx [m_h,\text{TeV}]$. The freeze-in production mechanism partially decouples the DM abundance from the active-sterile mixing $\theta$, thus complying with bounds from stability and indirect detection; moreover the resulting DM spectrum is ``colder'' with respect to the DW one, relaxing structure formation bounds.

Given that the ISS simultaneously accounts for neutrino masses and leptogenesis, or for neutrino masses and DM, it is natural to ask if a common solution to all these three problems can be achieved in this framework. Given the different mass scales involved in the BAU and DM production, one could investigate a split version of the (2,3) ISS, with the lighter pseudo-Dirac pair at the GeV scale accounting for leptogenesis via the ARS mechanism, and the heavier one at the TeV scale accounting for DM production via freeze-in decay. This scenario, however, is not successful in achieving the task~\cite{Abada:2017ieq}: viable DM solutions provide a BAU below the measured value, while viable BAU solutions overproduce DM through the DW mechanism, and are thus excluded in the (2,3) ISS. The (2,2) ISS, where there is no DM candidate, is nevertheless a viable scenario for BAU.

\section{Conclusion}
Sterile fermions can provide a common solution to the SM observational problems, namely the neutrino masses and the lepton mixing, the existence and properties of DM and the observed BAU. A minimal common solution for all the three problems is provided in the $\nu$MSM, although the requirement of simultaneously viable BAU and DM results in a quite fine-tuned scenario. The ISS provides an alternative mechanism to account for both neutrino physics and DM, or for neutrino physics and BAU, but BAU and DM solutions appear in different regions of the parameter space.

\end{document}